\documentclass[twocolumn,prl,amsmath,amssymb]{revtex4}
\usepackage[dvips]{graphicx}

\begin{document}

\def\lsim{\mathrel{\rlap{\lower4pt\hbox{\hskip1pt$\sim$}}
    \raise1pt\hbox{$<$}}}
\def\gsim{\mathrel{\rlap{\lower4pt\hbox{\hskip1pt$\sim$}}
    \raise1pt\hbox{$>$}}} 
\newcommand{\vev}[1]{ \left\langle {#1} \right\rangle }
\newcommand{\bra}[1]{ \langle {#1} | }
\newcommand{\ket}[1]{ | {#1} \rangle }
\newcommand{\ev}{ {\rm eV} }
\newcommand{\kev}{{\rm keV}}
\newcommand{\mev}{{\rm MeV}}
\newcommand{\gev}{{\rm GeV}}
\newcommand{\tev}{{\rm TeV}}
\newcommand{\mpl}{$M_{Pl}$}
\newcommand{\mw}{$M_{W}$}
\newcommand{\Ft}{F_{T}}
\newcommand{\Zparity}{\mathbb{Z}_2}
\newcommand{\BLambda}{\boldsymbol{\lambda}}
\newcommand{\be}{\begin{eqnarray}}
\newcommand{\ee}{\end{eqnarray}}
\newcommand{\cl}{95\% C.L.}

\def\rd{\mathrm{d}}
\def\lep{{\sc lep} }
\def\lepone{{\sc lep} 1 }
\def\leptwo{{\sc lep} 2 }
\def\aleph{{\sc aleph} }
\def\opal{{\sc opal} }
\def\delphi{ {\sc delphi} }
\newcommand{\muth}{ \mu_{\mathrm{th}} }
\newcommand{\shad}{ \sigma_\mathrm{had} }

\title{Constraining Light Colored Particles with Event Shapes}
\author{David E. Kaplan and Matthew D. Schwartz}
\affiliation{Department of Physics and Astronomy
Johns Hopkins University
Baltimore, MD 21218, U.S.A.}

\begin{abstract}
Using recently developed techniques for computing event shapes with Soft-Collinear Effective Theory, \lep 
event shape data is used
to derive strong model-independent bounds on new colored particles.
In the effective field theory computation, colored particles contribute in loops
not only to the running of $\alpha_s$ but also to the running of hard, jet and soft functions. Moreover, 
the differential distribution in the effective theory explicitly probes many energy scales,
so event shapes have strong sensitivity to new particle thresholds. Using thrust data from \aleph and
{\sc opal}, colored adjoint fermions (such as a gluino) below 51.0 GeV are ruled out to 95\%
confidence.
This is nearly an order-of-magnitude improvement over the
previous model-independent bound of 6.3 GeV.
\end{abstract}

\maketitle

Despite the fact that particle physics experiments have been running at and above 91 GeV center of
mass energies for over two decades, 
it is not known if the standard model represents the complete particle content below this scale. 
For particles which carry no standard model quantum numbers, 
the only hope of producing them at colliders is through the Higgs, if there is a Higgs,
and if they couple to it, or indirectly through off-shell intermediate states.
But surprisingly, even colored particles, which interact with the strong force, 
are not significantly constrained. 
As long as they have small or vanishing couplings to electroweak gauge bosons, 
current data allows mass ranges well below the weak scale. 
A good example is a color adjoint Majorana fermion,
such as the gluino in supersymmetric theories. 

The gluino is a color octet and thus should have a large production cross section at hadron colliders, 
a non-negligible contribution to four-jet events at {\sc lep}, 
and a significant effect on the running of $\alpha_s$.  
Most of the current bounds on a color octet fermion depend on how it hadronizes and how it decays. 
For example,
\begin{itemize}
\item If the gluino decays to two quarks and a very light neutralino, hadron collider data rules it out up to 308 
GeV at 95\% confidence level (C.L.)~\cite{:2007ww}.
 A recent study has shown that the Tevatron could probe gluino masses up to 150 GeV in the same decay channel
independent of the neutralino mass \cite{Alwall:2008ve}.
\item If the gluino is stable on detector lifetimes, 
\aleph has excluded masses lighter than 26.9 GeV~\cite{Heister:2003hc}.
\item A bound of 12 GeV, for a fixed $\alpha_s(m_Z) = 0.118$,
 has been set based on the gluino's potential contribution to the
parton distribution functions~\cite{Berger:2004mj}.  
A strict lower bound ({\it i.e.}, independent of $\alpha_s$) has not been set.
\end{itemize}

As for a model independent limit, \aleph  \cite{Barate:1997ha}
performed an analysis on four-jet observables as a measurement of the strong coupling constant and QCD color factors.
  The analysis found a good fit to QCD and ruled out gluinos below 6.3 GeV at \cl.
An independent study~\cite{Janot:2003cr}, which included the use of electroweak precision data,
arrived at the same lower limit.
%
%produced a similar result
%additional hadronic decay modes of the $Z$ (e.g. the $q\bar{q}\tilde{g}\tilde{g}$ final state). 
% This analysis arrives at the same scale, 6.3 GeV at \cl, independent 
%of how the gluino hadronizes or decays. 
In both cases, the bound comes essentially from the cross section
 for $q\bar{q}\tilde{g}\tilde{g}$ production which is very sensitive to the gluino mass at \lepone energies.
The scale 
6.3 GeV is where these searches lose sensitivity and can be taken as the current model-independent bound on the gluino mass. 

Besides real production, new colored particles can be seen indirectly by their virtual effects. For example, any particle with
color will contribute to the running of $\alpha_s$. Some of the current model-independent bounds come from fitting
the 1-loop $\beta$-function coefficient -- which is sensitive to the number of flavors, $n_f$ -- to values of $\alpha_s$ measured at different energies.
For example, {\delphi} has done a study of mean values of event shapes and other inclusive observables leading to 
$n_f = 4.7 \pm 1.2$ (and $n_f=4.75 \pm 0.44$ when combined with low energy thrust data) and ruling out gluinos less than 5 GeV.
This study includes data from 14 to 200 GeV. 
However, by averaging the observables -- for example, into
the mean thrust ${\overline T}(Q)$ -- this approach is not optimized to take full advantage of the available data.

A nominally positive feature of totally inclusive observables, such as ${\overline T}(Q)$, is that only one scale 
appears, so the perturbation series in $\alpha_s$ cannot be 
spoiled by the appearance of large logarithms. However, 
in searching for new particles through radiative corrections it is precisely these logarithms which have the most
valuable information. In order to trust a differential calculation where the logarithms are relevant, the logs must
be resummed. For many years resummation of event shapes was only available at next-to-leading order~\cite{Catani:1992ua},
which was insufficient to provide strong bounds on new physics because of large theoretical uncertainties.
Recently, however, the thrust distribution was resummed to next-to-next-to-next-to-leading logarithmic order using
techniques of effective field theory~\cite{thrust2}. Including matching to
 recent next-to-next-to-leading fixed-order (NNLO)
event shapes~\cite{GehrmannDeRidder:2007hr}, the theoretical uncertainty on the $\alpha_s$ extraction from \lep
was finally reduced to be sub-dominant to other uncertainties for the first time. 
Moreover, besides reducing the uncertainty, the effective theory approach
makes explicit that $\alpha_s$ is probed at many scales and so the 
sensitivity to new physics should be strong. Thus, it is natural to try to improve the model-independent bounds on
new colored states using these recent theoretical advances. In this letter, we use these insights to improve
the model-independent bound on the gluino by nearly an order of magnitude.

The thrust distribution was shown in~\cite{thrust1,thrust2} to have the form
\begin{equation}\label{Rtau}
R(\tau) =\frac{1}{\shad} \int_0^\tau\frac{\rd \sigma}{\rd \tau'} \rd \tau' 
=\frac{1}{\shad}\left[ R_2(\tau) + r(\tau)\right]
\end{equation}
where $\tau\equiv 1-T$.
Here, the matching function $r(\tau)$ is defined as the difference between the fixed-order
thrust distribution and the fixed-order expansion of the resummed distribution. We use $r(\tau)$
at next-to-next-to-leading order, {\it i.e.} to $\alpha_s^3$.

The function $R_2(\tau)$ in Eq.\eqref{Rtau} is the resummed distribution.
It can be calculated using Soft-Collinear Effective Theory~\cite{Bauer:2000yr,Bauer:2001yt,Beneke:2002ph}
with insights from 
from~\cite{Bosch:2004th,Becher:2006mr,Bauer:2006mk,Bauer:2006qp,Fleming:2007qr,Becher:2006qw}.
The result is~\cite{thrust2}:
\begin{multline}\label{scet}
  R_2(\tau) = 
\exp\left[ 4 S (\mu_h, \mu)  - 2 A_H (\mu_h, \mu) - 8 S (\mu_j, \mu)\right. \\
\left.+ 4 A_J (\mu_j, \mu) + 4 S (\mu_s,\mu) + 2 A_S (\mu_s, \mu) \right] \\
  \times H(Q,\mu_h) \left[ \widetilde{j} (\partial_{\eta},\mu_j)\right]^2 
\widetilde{s}_T (\partial_{\eta},\mu_s) \left[
 \frac{e^{- \gamma_E\eta}}{\Gamma ( \eta +1)} \right] \,.
\end{multline}
The derivatives in Eq.~\eqref{scet} are to be taken analytically and then $\eta$ set to
its canonical value $\eta=4 A_\Gamma(\mu_j,\mu_s)$. 
Here, $H(Q,\mu)$ is the hard function and $\widetilde{j}(L,\mu)$
and $\widetilde{s}(L,\mu)$ are the Laplace transforms of the jet and soft functions; all of these have power
series expansions. $S(\nu,\mu)$ and $A_X (\nu,\mu)$
are auxiliary functions defined as integrals over various anomalous dimensions. Explicit expressions for these
functions can be found in~\cite{thrust2}. 
The scale $\mu$ in Eq.\eqref{scet} is arbitrary and the distribution is formally independent of it, but different values
of $\mu$ can be chosen for calculational convenience.

The formula~\eqref{scet} is a
simplified version of the one in~\cite{thrust2}, valid when the scales are set to their canonical
values:
\begin{equation}
\mu_h = Q,\quad \mu_j= Q \sqrt{\tau},\quad \mu_s = Q \tau
\end{equation}
As mentioned above, a calculation of mean thrust, or a fixed-order calculation of differential 
thrust, would only probe $\alpha_s(Q)$ at a single scale, the hard scale $\mu_h = Q$.
But the differential thrust distribution probes even lower scales. For example,
in the two-jet limit thrust reduces to the sum of hemisphere masses, $Q^2 \tau \sim M_L^2 + M_R^2 $. 
The effective
theory expression associates this mass scale with the scale of jet functions, and probes it through 
$\mu_j \sim Q\sqrt{\tau}$. Actually, the effective theory makes it apparent that even lower scales, associated with
soft modes of QCD, are relevant. These are probed by the soft scale
$\mu_s\sim \mu_j^2/\mu_h \sim Q\tau$, which is a type of seesaw scale lower than both of the physical
external scales $Q$ and $Q\sqrt{\tau}$~\cite{thrust1}. Since $\alpha_s$ is larger at lower energy,
resumming logs of the soft scale is critical to generating an accurate thrust distribution.

Because the differential thrust distribution is sensitive to many scales, 
it would be sensitive to the presence of new colored particles with a variety of masses. These
new states would affect the running of $\alpha_s$, through the QCD beta function, as well as
the hard, jet, soft anomalous dimensions -- 
which appear implicitly in~\eqref{scet} --
and the fixed-order hard, jet, and soft functions, $H,\widetilde{j}$ and $\widetilde{s}$. 

Throughout the following
we modify the standard model by adding $\Delta n_f$ new flavors of mass $m$ 
at a threshold scale $\muth$. For example, a new massive 
quark corresponds to $\Delta n_f=1$ 
and a gluino to $\Delta n_f=3$~\cite{foot0}.
Below the scale $\muth$, the new flavors are integrated
out, insuring decoupling as $m\to\infty$. This will, in general, induce discontinuities in $\alpha_s(\mu)$
and in the hard, jet and soft functions, all of which are unphysical by themselves. The resulting thrust distribution, however,
must be smooth. In fact, one can show
 that the effective field theory distribution
 is independent of $\muth$ order-by-order in perturbation theory~\cite{Rodrigo:1993hc}.
For simplicity, we take $\muth=m$ and match $\alpha_s$ at one-loop.
To avoid having to run the jet and soft functions through the threshold, we choose $\mu=m$ in Eq.\eqref{scet}
when $\mu_s<m<\mu_j$. For $m<\mu_s$, we take $\mu=\mu_s$ and for $m>\mu_j$ we take $\mu=\mu_j$. 

To demonstrate the sensitivity to new states, we
begin by looking at a single data set, the \aleph data from \lepone at 91.2 GeV~\cite{aleph1}.
We perform a bin-by-bin 
correction for hadronization and quark masses using {\sc pythia}
{\tt v.6.409}~\cite{Sjostrand:2006za}.
Using the fit region
$0.10 \le \tau \le 0.24$ the soft scale $\mu_s$ probes 9 - 22 GeV and the
 jet scale $\mu_j$ probes 29 - 45 GeV.
Thus, if there are $n_f$ flavors below 9 GeV, we do not have to worry about an explicit threshold and
may simply run $\alpha_s$ and the other objects using this value of $n_f$ throughout 
(a more refined procedure is described below). To derive
 a bound on the number of light flavors, we perform a least-squares fit
to the experimental data. For the errors used in the fit, we include both the 
experimental statistical uncertainty and also the statistical uncertainty in the
fixed-order thrust distribution. The fixed-order 
result was calculated numerically, with somewhat slow convergence at NNLO,
and to be conservative we rescale the NNLO uncertainties by a factor of 1.5 to 
account for the fact that the errors may have been underestimated.
A combined fit with two free parameters gives $\alpha_s(m_Z) = 0.1169 \pm 0.0004$
and $n_f = 5.32 \pm 0.59$, where the errors are statistical only.
%This by itself rules out gluinos below 9.1 GeV, but we will derive
% a stronger bound below.

\begin{figure}
\resizebox{\hsize}{!}{\includegraphics{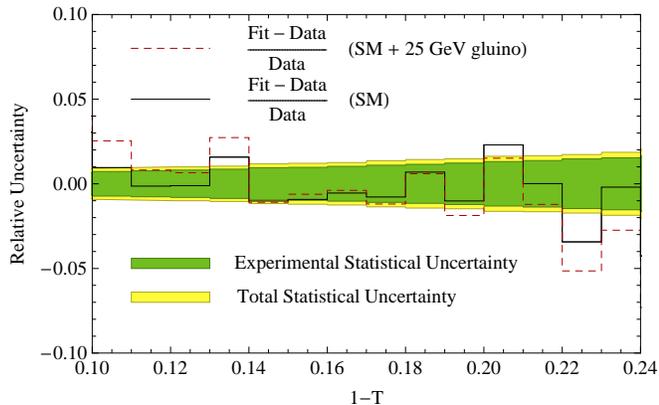}}
\caption{Theoretical prediction versus \aleph data at \lepone
 for the standard model and the standard model with a 25 GeV gluino.
The total statistical uncertainty band includes theoretical statistical
uncertainty from the Monte Carlo used to generate the NNLO fixed-order
thrust distribution.}
\label{fig:errorplot}
\end{figure}

For a second example, 
using the same \lepone data set, we note that a gluino of mass $m=25$ GeV lies
outside the range of scales probed by the hard, jet, and soft functions. 
Thus, it can be modeled by taking $\Delta n_f=3$ for the hard and jet functions,
and $\Delta n_f=0$ for the soft function.
Performing the fit with these values, we find $\chi^2 = 31.7$ with the gluino compared
to $\chi^2=11.9$ for the standard model, with 13 degrees of freedom. The fits for the two models
 are shown in Figure~\ref{fig:errorplot}, where
it is clear that the model with the gluino is systematically worse.

To properly scan over masses, we must specify how to handle the thresholds.
First, consider the total hadronic cross section, $\shad$.
The exact leading order dependence of $\shad$ on the new particle mass 
can be extracted from~\cite{Hoang:1994it}. For $m<\mu$, the contribution to the total cross section is 
proportional to $\Delta \shad = \alpha_s^2(\mu)\left(\rho_V(\frac{m^2}{Q^2}) + \rho_R(\frac{m^2}{Q^2}) 
+ \frac{1}{4}\log(\frac{m^2}{\mu^2})\right)$, where $\rho_V$ is the virtual contribution which vanishes
at $m=\infty$ and $\rho_R$ is the real emission contribution which vanishes for $m>Q/2$.
The explicit log compensates the $\mu$-dependence of $\alpha_s$ and 
is necessary to have a smooth $m\to0$ limit. 
We will use this exact expression $\Delta \shad$ 
for the new physics contribution to $\shad$ in Eq.~\eqref{Rtau},
but observe that, as shown in~\cite{Hoang:1994it}, 
it is well approximated for $0<m<Q$ by the leading power in $m^2/Q^2$. 
Actually, it is not clear whether the experiments would have included
decay products of real gluinos in their event selection for the thrust distribution,
so in the spirit of providing a model-independent bound, we allow $\Delta \shad$ to
scan between 0 and the cross section for $\Delta n_f$ additional massless flavors. This
variation is included in the uncertainty band described below.

\begin{figure}[t]
\resizebox{\hsize}{!}{\includegraphics{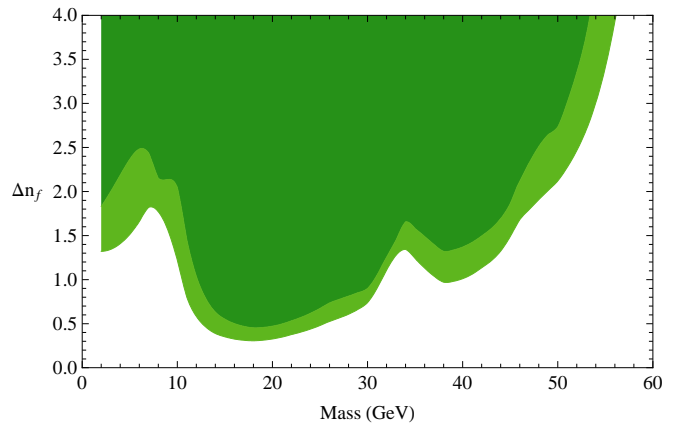}}
\caption{Bounds on light colored particles from \lep data. The darker region is completely
excluded at 95\% confidence. The lighter region is an uncertainty band including
estimates of various theoretical uncertainties.}
\label{fig:contplot}
\end{figure}

The exact contributions of massive
colored states to the jet, soft, and hard functions are not known, 
but since the same loops and real-emission diagrams are
relevant for them as for $\Delta\shad$, 
it is likely that the result would be similar to that of $\Delta \shad$. 
Thus, we assume the leading power is linear in $m^2/\mu_h^2$ for the hard function, 
$m^2/\mu_j^2$ for the jet function, and $m^2/\mu_s^2$ for the soft function.
That is, we take $H, \widetilde{j}$ and $\widetilde{s}$ to interpolate
between the expression for $n_f=5+\Delta n_f$ flavors at $m=0$ and
$n_f=5$ flavors at the relevant threshold. 
This removes any remaining discontinuity in the thrust distribution, and should be
a good approximation to the (unknown) exact result. 
In a similar vein, the matching correction, $r(\tau)$ in Eq.~\eqref{scet}, formally takes
place at the hard scale $Q$. However, it depends on $n_f$ and would be
discontinuous as $m$ crosses $Q$ unless 
the discontinuity is removed by inclusion of explicit mass corrections. 
We use an interpolation also linear in $m^2/Q^2$
for this effect.
Using this model for the mass thresholds, in lieu of the exact result,
introduces some theoretical uncertainty. To account for
that uncertainty, we explore some variations of the model and include the errors in our final 
bound, as described below.

With this treatment of the threshold effects, 
the thrust distribution is smooth and can be compared with the data for each
$m$ and $\Delta n_f$. We perform a combined fit to
the {\sc aleph}~\cite{aleph1} and {\sc opal}~\cite{opal1,opal93} data sets from 
$91.2 - 206$ GeV~\cite{foot1,foot2}.
The fit regions used are $0.1<\tau<0.24$ for \lepone, 
and $0.04<\tau<0.25$ for \aleph \leptwo and $0.05<\tau<0.22$ for \opal \leptwo.
The data are corrected bin-by-bin for hadronization and bottom/charm mass effects using {\sc pythia}.
We perform a least-squares fit of the theoretical prediction to the corrected data,
using errors which include both the experimental statistical errors and the statistical errors
of the NNLO fixed-order calculation, rescaled by 1.5, as described above. For the standard model,
the $\chi^2$ is 85.7 for 78 degrees-of-freedom. For each value of $m$ and $\Delta n_f$, we
minimize $\chi^2$ and compute the maximum likelihood ratio as compared with the standard model.
 The resulting \cl~bound is shown in Figure~\ref{fig:contplot}. For $\Delta n_f=3$, the limit is
$m_{\widetilde g} > 52.5$ GeV. For a real gluino (with the appropriate group theory factors differing
from $\Delta n_f=3$ at higher orders), the bound differs by 0.03 GeV.

To account for the theoretical uncertainty, we include an uncertainty band 
(the light shaded region in Figure~\ref{fig:contplot}). This subsumes the following variations: 
({\it i}) Removing the lowest bins from each data set in the fit.
({\it ii}) Not interpolating the total cross-section and matching correction 
(we include variations  both with $n_f=5$ and with $n_f=5+\Delta n_f$ in $\shad$ and $r(\tau)$).
({\it iii}) Varying parameters in the hadronization model between {\sc pythia}'s
default values and the \aleph \cite{Barate:1996fi} and \opal \cite{Biebel:1996mc} optimized tunings 
({\tt PARJ(81) = 0.290, 0.292, 0.250} and {\tt PARJ(82)=1.00, 1.57, 1.90}).
({\it iv}) Using a power correction proportional to $m/\mu_x$ instead of $m^2/\mu^2_x$ for the threshold corrections.
The band in Figure~\ref{fig:contplot} includes the maximal and minimal bounds 
at \cl~for each value of $m$ and $\Delta n_f$. For the gluino, this gives $m_{\widetilde g} > 51.0-54.0$ GeV.
For our final bound, we take the least restrictive value, $m_{\widetilde g} > 51.0$ GeV.

From Figure~\ref{fig:contplot}, it is clear that this method has the strongest sensitivity in an intermediate
mass range, 10 GeV $\lsim m \lsim$ 40 GeV. This range roughly coincides with the scales probed 
by the jet and soft functions in the fit regions of the thrust distributions. For masses below about 10 GeV,
the mass threshold lies outside of the fit regions and the effect on the event shape can be
partially compensated by a change in $\alpha_s$. When the mass falls inside the range of the thrust
distribution, it is more difficult to compensate by rescaling $\alpha_s$, hence the stronger bound.
With additional independent constraints on $\alpha_s$, for example, from the lattice~\cite{Mason:2005zx}
or from $\tau$ decays~\cite{Davier:2008sk}, one might be able to close the light mass window more tightly. 
This might, for example, even rule out additional light colored triplets or scalar adjoints.
However, as the lattice and $\tau$-decay determinations of $\alpha_s$ (which take place
at similar scales) are themselves inconsistent by more than two standard deviations,
it is unclear whether a definitive bound could be obtained in this way. The main result of this paper
is that event shapes alone are sufficient to exclude
light and intermediate mass gluinos up to 51 GeV, independently of their decays.

The authors would like to thank Thomas Becher, Bill Gary, Steve Mrenna and 
Morris Swartz for helpful conversations. 
This work is supported in part by the National Science Foundation under grant NSF-PHY-0401513, the Department of Energy's OJI program under grant DE-FG02-03ER4127, and the Johns Hopkins Theoretical Interdisciplinary Physics and Astronomy Center.


\begin{thebibliography}{1}
\bibitem{foot0}
A gluino is only equivalent to $n_f=3$ flavors in the leading $n_f$ dependence. However, 
for the mass bound the difference between a gluino and $\Delta n_f=3$ is negligible.
\bibitem{foot1}
The \aleph data at 200 GeV are not included because the $\chi^2$ of its fit is
 unacceptably large $\sim 24$ for 4 d.o.f.. The effect
of omitting this outlier falls well within the uncertainty band
for the final mass bounds, 
but restores the total $\chi^2$ to $\sim 1$/d.o.f..
The rest of the \aleph data, up to 206 GeV, are included in the fit.
\bibitem{foot2}
The \opal \lepone data is taken from~\cite{opal93} instead of ~\cite{opal1} because the data
from~\cite{opal93} 
is more finely binned and thus more sensitive to changes in the 
event shape.
\end{thebibliography}
\end{document}